\documentstyle[twocolumn,aps,prl]{revtex}

\begin{document}
\draft
\title{
Bloch Electron on a Triangular
Lattice and Quantum Ising Model in a Transverse Field }

\author{ Jukka A. Ketoja}
\address{ Department of Physics, {\AA}bo Akademi,\\
Porthansgatan 3, FIN-20500 {\AA}bo, Finland}
\author{ Indubala I. Satija}
\address{
 Department of Physics and\\
 Institute of Computational Sciences and Informatics,\\
 George Mason University,\\
 Fairfax, VA 22030}

\date{\today}
\maketitle
\begin{abstract}
The tight binding model for an electron on an
 anisotropic triangular lattice in a uniform magnetic field
is studied using a decimation scheme. The model exhibits a transition
from critical to localized phase and the phase diagram is described
in terms of three nontrivial renormalization fixed points for the band
edges. These subcritical, critical, and supercritical universality
classes also describe
the corresponding states of the quantum Ising chain in a modulating transverse
field. The only exception is the
conformally invariant point of the Ising model which has no analog
in the electron problem.
\end{abstract}

\pacs{75.30.Kz, 64.60.Ak, 64.60.Fr}
\narrowtext

The problem of a two-dimensional Bloch electron in an irrational magnetic
flux is important both theoretically and experimentally.\cite{Sok}
In this paper, we study the electron on a
triangular lattice with anisotropic
couplings between the nearest neighbor sites.
We use a renormalization approach to obtain the phase diagram and
universal scaling properties of the model. The results are compared with
those of the Ising model in a transverse modulating field
of periodicity incommensurate with the periodicity of the spin chain.
With the exception of conformally invariant
 point\cite{Hankel} of the Ising model\cite{KS},
the band edge as well as the band center states of the two models
belong to the same universality class.

In the tight binding approximation, the energy spectrum for a single Bloch
band of an anisotropic triangular lattice with couplings $t_1$
and $t_2$ is\cite{Wannier},
\begin{eqnarray}
\epsilon(k_x,k_y)&=&-t_1 \cos[{a\over 2} (\sqrt3 k_x-k_y)]\nonumber\\
&-&t_1 \cos[{a\over 2} (\sqrt3 k_x+ k_y)] +t_2 \cos(ak_y)
\end{eqnarray}
where $a$ is the lattice spacing.
Using Peierl's substitution $\hbar {\bf k} \rightarrow {\bf p} -{e
\over c} {\bf A}$ with ${\bf p} = {\hbar \over i} \nabla$ and
the Landau gauge ${\bf A} = ( 0, x, 0 ) H$ for the vector potential
{\bf A},
the two-dimensional problem is reduced to the one-dimensional
tight binding model (TBM)  \cite{Wannier}
\begin{eqnarray}
& & \cos[ \pi (i\sigma +(\phi+\frac{\sigma}{2}))] \psi_{i+1}
+ \cos[ \pi (i\sigma +(\phi-\frac{\sigma}{2}))] \psi_{i-1}
\nonumber\\
&+& \lambda \cos[2 \pi (i\sigma +\phi)] \psi_i = E \psi_i
\end{eqnarray}
which we call the triangular model.
Here, $i=2x/(a\sqrt3)$ and $\sigma$ is the dimensionless magnetic
flux in units of Bohr magneton, $\sigma=(eH/hc)a^2$
and $\lambda=t_1/t_2$. The isotropic triangular lattice case $t_1=t_2$
was
studied by Claro and Wannier \cite{Wannier} where the model was found to
exhibit
the hierarchial structure in the energy spectrum similar to that of the square
lattice case studied by Hofstadter \cite{Hoft}.

In this paper, we show that as a function of $\lambda$, the model
exhibits a phase transition from critical (C) to localized
(L) states. This is in contrast with the case of the square lattice
where the anisotropic model exhibits a transition from
extended (E) to localized states. Therefore, unlike the square lattice case,
the model does not exhibit the E phase of KAM type and hence the
weak coupling limit of the model is nontrivial. Furthermore, the critical
phase exists in a finite parameter interval.
We use our recently developed decimation method \cite{KS,Ketoja} to
study the scaling properties in various parameter ranges.
At the band edges, the universality classes of the model
are described by the following three non-trival fixed points of
the renormalization operator
(the term "fixed point" is used also for periodic cycles
of the renormalization operator as the latter are fixed points
of an higher-order iteration of the operator):
1) the weak coupling (subcritical) fixed point
 ($\lambda < 1$), 2)
the critical fixed point corresponding to $\lambda=1$ describing the
onset of the C-L transition, and 3) the strong coupling (supercritical)
fixed point ($\lambda > 1$)
describing the fluctuations in the exponential decay in a localized
wave function \cite{KSloc}.

The C-L transition and the universal properties
of the triangular model (2) are compared with those of
the quantum Ising model
(QIM) in a transverse field $h_i$ for which the Hamiltonian reads
\begin{equation}
H=-\sum_i [ \sigma^x_i \sigma^x_{i+1} + 2 h_i \sigma^z_i ].
\end{equation}
Using the methods described by Lieb et al. \cite{Lieb}, the eigenvalue
equation for the
spin problem can be written in a TBM form, which for
$h_i = \lambda \cos [\pi (i\sigma +\phi)]$ becomes
\begin{eqnarray}
& & \cos[ \pi (i\sigma +\phi)] \psi_{i+1}
+ \cos[ \pi (i \sigma+ \phi)] \psi_{i-1}
\nonumber\\
&+& \lambda \cos[2 \pi (i\sigma +\phi)] \psi_i = E\psi_i
\end{eqnarray}
Here $E=-\lambda+(\bar{E}^2 /4-1)/(2 \lambda)$, where $\bar{E}$ is the energy
of the Ising model (3).\cite{IIS}

The QIM was recently studied for $\sigma$ equal to the inverse golden
mean $\sigma=(\sqrt5-1)/2$.
The model exhibits critical states for $\lambda \leq 1$ where
$\lambda=1$ is the localization transition point.
In the QIM, this transition has an additional significance as it also
corresponds to the
magnetic transition to the long range order. At this point,
the lowest energy quantum state is believed to be conformally
invariant.\cite{Hankel}  Our recent decimation studies \cite{KS}
confirmed this phase diagram of the model and showed that the subcritical
regime was described by a unique fixed point of the renormalization flow.
However, at the onset of localization, the renormalization flow was
attracted by a different fixed point resulting
in a different universality class.

It is rather interesting to note the similarities between
these two seemingly unrelated
problems. Both are described by a nearest-neighbor
 TBM where the diagonal as well as the
off-diagonal
terms are modulating: the periodicity of the diagonal term is twice the
periodicity of the off-diagonal term. The only
 difference between these two
models is the fact that unlike the QIM, the triangular model
shows a relative phase difference between the diagonal and off-diagonal
term which is $\sigma$ dependent.

We applied the renormalization approach\cite{KS} to the triangular model.
Our main focuss is to
compare the phase diagram and the universal properties of the triangular model
with those of the QIM.
The TBM is written in the following form, where all sites of the lattice
except those labelled by the Fibonacci numbers $F_n$ are decimated:
\begin{equation}
f_n(i) \psi(i+F_{n+1})=\psi(i+F_n) + e_n(i) \psi(i).
\end{equation}
The additive property $F_{n+1} =F_n + F_{n-1}$ provides exact recursion
relations for the decimation functions $e_n$ and $f_n$:\cite{KS,KSloc}
\begin{eqnarray}
e_{n+1} (i)= - {A e_n (i) \over 1+Af_n (i)} \\
f_{n+1} (i)= {f_{n-1} (i+F_n) f_n(i+F_n)\over 1+Af_n(i)} \\
A = e_{n-1} (i+F_n) + f_{n-1} (i+F_n)e_n(i+F_n). \nonumber
\end{eqnarray}
In this renormalization approach,
the C phase manifests itself in a nontrivial limit cycle
of period $p$, which is typically a multiple of $3$,
for the absolute values of decimation functions $e_n$ and $f_n$
as $n \to \infty$.  This results in nontrivial scaling ratios
\begin{equation}
\zeta_i = \lim_{n \rightarrow \infty} |\psi(F_{i+pn})/\psi(0)|.
\end{equation}
Here it is assumed that the phase $\phi$ is chosen so that
the main peak of the wave function lies at $i=0$.
In contrast, the scaling
ratios approach unity in the E phase.

In the L phase, we write the wave function as\cite{KSloc},
\begin{equation}
\psi_i = e^{ -\gamma |i|} \eta_i
\end{equation}
where $\gamma$ is the Lyapunov exponent which vanishes in the E and C
phase. $\eta$ describes the fluctuations
in the exponentially decaying part of the wave function.
The spatial dependence of $\eta$ is given by a TBM\cite{KSloc} which
is studied by the decimation scheme described above.
For the triangular lattice, $\gamma = ln(\lambda)$ \cite{Wannier}.
Due to lack of duality, an analogous expression has not been derived
analytically for the QIM. However, a numerical computation
of the Lyapunov exponent\cite{thank} is found to be in agreement
with this formula.

Detailed study of the renormalization flow
shows that for the triangular model (2),
the decimation functions approach asymptotically the same $3$-cycle
for both the upper and the lower band edge (maximum and minimum energy).
For the QIM, this identity between
the scaling properties of the upper and lower band edge
is true only in the subcritical and supercritical regimes
but not at the critical point (see next paragraph).
Furthermore, in both subcritical and supercritical regime, the triangular
model and the QIM are described by the
same universality class, i.e. the scaling ratios $\zeta$
for both models
are identical. In the triangular model the wave function has
exact symmetry at the band edges whereas in the QIM
the wave function is only asymptotically symmetric.

At the critical point $\lambda =1$ corresponding to the onset of
localization, the universal characteristics of the triangular lattice
band edges are the same as those of the upper band edge of the QIM.
However, the lower band edge of the QIM has zero energy
and the corresponding state
is believed to be conformally invariant. Unlike the
upper band edge $3$-cycle, the renormalization flow for
the decimation functions at $\bar{E}=0$
converges to a period-$1$ fixed point. In addition,
the wave function at the conformal point is asymmetrical
vanishing on one side of main peak.\cite{KS}
 The "conformal" renormalization
fixed point of the QIM does not map
to any quantum state of the triangular model.

Fig. 1 shows the self-similar wave functions
in the three universality classes for the triangular model
at the upper band edge. They also describe the asymptotic wave functions
at the upper band edge of the QIM.
The dominant peaks in the wave functions are labelled by the
rational approximants of $\sigma^3$ and its harmonics.\cite{KS}
The height of these peaks with respect to the central peak define universal
scaling ratios.
For example, the universal scaling ratios of
the peaks at three successive Fibonacci numbers
in the subcritical, critical, and supercritical
 phases are ($0.825, 0.908, 0.836$),
($0.238, 0.303, 0.291$), and ($0.267,0.311,0.121$), respectively (see
Fig. 1).
These scaling ratios are
identical to the corresponding scaling ratios at the lower band edge
with the exception of the
the conformal point in the QIM
characterized by the
fixed point scaling $0.415$.

The study of the band center indicates how the relative phase
difference in the off-diagonal terms of the triangular model and the QIM
can lead to subtle symmetry properties in the asymptotic
renormalization dynamics of the two models. At the band center with
$\lambda=1$,
the wave functions are asymmetric in both systems leading to
different 6-cycles and scaling ratios (also different
from those characterizing the band edges) on the positive and negative side
of the lattice.\cite{KSC} The scaling properties on the positive
(negative) side of the triangular model are asymptotically the same as
those of the QIM on the negative (positive) side.
However, the correspondence is seen only with the shift of three
decimation levels between the two systems.
Moreover,  the critical value of the phase $\phi$ setting the main
peak at the origin is different for the models. In the QIM
it is $1/4$ (the same as at the lower band edge) as for the critical model
the critical phase is $3/8$
which is the arithmetic mean of the critical phases at the upper and lower
band edges.

The Harper equation, which describes Bloch electrons on a square
lattice, also models the isotropic XY model in a quasiperiodic field.
\cite{KS}
It is rather interesting that the problem for the
Bloch electrons on a triangular lattice is related to the
fully anisotropic Ising limit of the XY model.
Although the correspondence between the TBM for the square lattice
 and the
isotropic XY model is exact, relationship between the triangular lattice
and the Ising model is through the asymptotic
universal scaling properties of the models.

Another interesting aspect of our studies is that
unlike the periodic case, details of the lattice
is relevant in determining the phase diagram and
universality classes in incommensurate
systems. For the square lattice, there is
a phase transition from E to L phase while in the case of a triangular
 lattice, the C-L transition is observed instead.
Therefore, for the square lattice, the weak coupling
fixed point is trivial while for the triangular case, the weak coupling
is described by the nontrivial fixed point and scaling ratios.
Furthermore, even in the L phase,
the scaling properties of the self-similar fluctuations
in the square and the triangular lattices are different.

The crossover from E-L to C-L phase diagram, as we change from the square to
the triangular lattice, can be understood due to the relationship between
the 2-dimensional Bloch and spin problem. Previous studies\cite{IIS}
 have shown
that in the presence of anisotropy, the E and the L phases
of the isotropic XY model are mediated by a critical phase with
self-similar characteristics. In the Ising limit, the width of the
E phase shrinks to zero resulting in the C-L phase diagram.
We like to mention that although the Harper equation with
next-nearest-neighbor interaction\cite{KSC,Thou} provides a means
to study the cross-over from square to triangular lattice,
understanding the E-L to C-L behavior is lot more complicated.

The results of the triangular model may be of relevance in experiments on
quantum dots.\cite{expt} Furthermore, since the one-dimensional QIM also
describes the two-dimensional classical Ising model with modulating exchange
along one direction, it could be realized in experiments
on magnetic superlattices. Our studies provide interesting mappings
between the thermodynamical properties such as the specific heat and
the susceptibility between these two models.

\acknowledgements

The research of IIS is supported by a grant from National Science
Foundation DMR~093296.
IIS would like to acknowledge the hospitality of National Institute
of Standard and Technology where part of this work is done.
JAK has been supported financially by the Academy of Finland
under contract 4385 and by the Niilo Helander Foundation.
JAK is also grateful for the hospitality of the Research Institute
for Theoretical Physics in Helsinki.

\begin{figure}
\caption{The wave function in the subcritical (a), critical (b),
and in the supercritical (c) triangular model for the golden mean
incommensurability at the upper band egde with $\phi=0$.
Note that in the supercritical case the figure descibes
the fluctuations $|\eta_i |$ and not
$|\psi_i |$.}
\label{fig1}
\end{figure}

\end{document}